\documentclass[prd,amsmath,aps,amssymb,nofootinbib,superscriptaddress,floatfix,twocolumn]{revtex4-2}
\usepackage{bm,amssymb,amsmath,siunitx,color,booktabs}
\usepackage[justification=raggedright,singlelinecheck=true]{caption}
\usepackage[mathscr]{euscript} 
\usepackage{graphicx}
\usepackage{upgreek}
\usepackage{dcolumn}
\usepackage{multirow}
\usepackage{bm}
\usepackage[colorlinks=true,linkcolor=red,urlcolor=blue,citecolor=blue]{hyperref}
\usepackage{url}
\usepackage{float}
\usepackage{caption}
\usepackage{subcaption}
\usepackage{makecell}
\usepackage{amsmath}
\usepackage{booktabs}
\usepackage{listings}
\usepackage{threeparttable}

\begin{document}

\title{Source Confusion of Massive Black Hole Binaries for the Taiji Mission}

\author{Qing Diao}
\affiliation{International Centre for Theoretical Physics Asia-Pacific, University of Chinese Academy of Sciences, 100190 Beijing, China}
\affiliation{Center for Gravitational Wave Experiment, National Microgravity Laboratory, Institute of Mechanics, Chinese Academy of Sciences, Beijing 100190, China}

\author{Hongxin Wang}
\affiliation{School of Software, Northwestern Polytechnical University, Xian 710072, China}

\author{Manjia Liang}
\affiliation{Center for Gravitational Wave Experiment, National Microgravity Laboratory, Institute of Mechanics, Chinese Academy of Sciences, Beijing 100190, China}

\author{He Wang}
\affiliation{International Centre for Theoretical Physics Asia-Pacific, University of Chinese Academy of Sciences, 100190 Beijing, China}

\author{Ziren Luo}
 \affiliation{Center for Gravitational Wave Experiment, National Microgravity Laboratory, Institute of Mechanics, Chinese Academy of Sciences, Beijing 100190, China}

\author{Minghui Du}
 \email{duminghui@imech.ac.cn}
 \affiliation{Center for Gravitational Wave Experiment, National Microgravity Laboratory, Institute of Mechanics, Chinese Academy of Sciences, Beijing 100190, China}

 \author{Peng Xu}
 \email{xupeng@imech.ac.cn}
 \affiliation{Center for Gravitational Wave Experiment, National Microgravity Laboratory, Institute of Mechanics, Chinese Academy of Sciences, Beijing 100190, China}
 \affiliation{Hangzhou Institute for Advanced Study, University of Chinese Academy of Sciences, Hangzhou 310024, China}

\begin{abstract}

We systematically investigate the source confusion of massive black hole binaries (MBHBs) for the  Taiji space-based gravitational wave mission. Source confusion, arising from the overlap of signals in both time and frequency domains, can degrade parameter recovery accuracy.
To assess this effect, we simulate three representative models MBHB populations to estimate event overlap events.
Assuming 100 detections per year, only 0.31-4.2 overlaps are expected annually.
Based on Fisher information matrix with the \texttt{IMRPhenomD} and \texttt{IMRPhenomHM} waveform models, we find that overlap significantly enlarges parameter uncertainties, while the inclusion of higher-order modes (HMs) effectively mitigates this effect. 
Severe confusion ($\Delta \mathcal{M}_z / \mathcal{M}_z<$ 0.2\%) occurs in fewer than 0.14\% across the three population models. 
The full Bayesian analysis further corroborates the Fisher predictions, and also reveals that HMs help break key parameter degeneracies, with or without signal overlap.  
These findings underscore the importance of incorporating HMs for accurate inference in future space-based observations. 
\end{abstract}

\maketitle

\section{Introduction}
The successful detection of gravitational waves (GWs) from the merger of stellar-mass black hole binaries GW150914 by the Laser Interferometer Gravitational-Wave Observatory (LIGO)~\cite{ligo_prl} marked the advent of gravitational-wave astronomy and established GWs as a novel probe of the universe. 
Due to the limitations imposed by seismic noise and Newtonian noise, the detection of millihertz-band GWs is expected to be achieved by space-based missions.  
In this context, the Taiji program in space, proposed by the Chinese Academy of Sciences, aims to deploy a triangular satellite constellation with $3\times10^6$ km arm lengths in a heliocentric Earth-like orbit, enabling precise detection of mid- and low-frequency ($10^{-4}-1$ Hz) GWs~\cite{HWRtaiji,2016NaturTaiji}. 
Compared to ground-based detectors, space-borne antennas offer access to the previously inaccessible frequencies~\cite{groundandspace}, and allow observation of a wide range of astrophysical and cosmological sources, including galactic white dwarf binaries, stellar-mass black hole binaries, extreme mass-ratio inspiral, massive black hole binaries (MBHBs), and potentially the relic stochastic GW background from the early universe. 
These signals are of great interest for cosmology, astrophysics, and fundamental physics~\cite{QFT_G,LISA_redbook2024,MBHB_SNR_testGR,christensen2018stochastic,EMRI}.

As space-based missions progress toward greater sensitivities, expanded frequency ranges, and extended life-times, signal overlaps in both the time and frequency domains are anticipated to become more prevalent~\cite{Li_2025}, resulting in source confusion and presenting challenges for accurate parameter estimation~\cite{Babak_2010_datachallenge,TDC_ren,du_TDC}. Previous studies have investigated the effects of signal overlap on parameter estimation, encompassing various detector architectures, from existing and next-generation ground-based detectors~\cite{ground-based} (like Advanced LIGO~\cite{adLIGO}, Cosmic Explorer~\cite{CE}, and the Einstein Telescope~\cite{ET}) to the space-based LISA mission~\cite{LISA,LISA_redbook2024}, and also covered a wide range of compact binary sources, including black hole binaries (BBHs), binary neutron stars (BNSs), and neutron star–black hole (NSBH) systems~\cite{ligo/virgo,10.1093/mnras/stab2358ET/LISA,Reali2022TheIO,johnson2024source,Reali2024IntermediatemassBH,bbh/tianqin/lisa,et/bbh/bns/nsbh,sourceR21,Janquart_2023,baka2025overlapping}. Qualitatively, the closer the coalescence times and frequency evolution tracks of two signals, the more severe the confusion effects. 
For instance, Wang \emph{et al.} ~\cite{Wangzm_2024} showed that a small difference in coalescence time is only a necessary condition for large deviations in parameter inference, and nearly identical chirp masses can lead to strong biases due to almost indistinguishable frequency evolutions. 

A key observation is that, as pointed out in Ref.~\cite{johnson2024source}, source confusion arises when signals overlap in both the time and frequency domains (\emph{i.e.} intersecting time-frequency tracks), rather than in either domain alone. 
The overlap in both domains is recognized as a key contributor to the deterioration of parameter estimation accuracy ~\cite{Wangzm_2024,johnson2024source}. 
Moreover, failing to model these overlaps can result in severe systematic biases for astrophysical inferences~\cite{Janquart_2023}. 

Among various targets of space-based GW detection, the detection and analysis of MBHBs not only lay the foundation for gravitational-wave astronomy involving supermassive black holes, but also offer unique insights into galaxy formation, black hole growth, and the nature of strong-field gravity\cite{LISA_redbook2024,galaxies12020017,testGR,colpi2019astro2020sciencewhitepaper,MBHB_SNR_testGR,MBHBformation}. MBHB signals, with masses typically within $10^4$–$10^7 M_\odot$,  fall squarely within the sensitivity bands of Taiji and LISA~\cite{massMBHB}. Some of these signals are expected to exhibit exceptionally high signal-to-noise ratios (SNRs), making them powerful probes for high-precision astrophysics and cosmology~\cite{MBHB_SNR_testGR}. 

However, the predicted merger rate of MBHBs remains highly uncertain, ranging from a few to several hundred events per year, depending on the underlying population synthesis model. 
Notably, several observational constraints and theoretical studies support the plausibility of high event rates (up to $\sim100$ per year)~\cite{LISA_redbook2024,ptaLISAdetectionsMBHB,Wang2024ai,massMBHB}. If realized, such high event rate implies that, during long-duration observations, MBHB signals may overlap in time and frequency, leading to source confusion and introducing systematic biases in parameter estimation. 
As crucial steps towards understanding and mitigating these effects, it is essential to first estimate the rate of source confusion under different population models, 
and quantify the impact of source confusion on parameter estimation. 

In this study, we consider three representative population models (\textsc{PopIII}, \textsc{Q3d} and \textsc{Q3nod}) to systematically assess the scenario of source confusion~\cite{population1,population2}. 
Based on these predictions, we apply the Fisher information matrix (FIM) framework to estimate how confusion from overlapping signals degrades parameter estimation precision. 
As an efficient approximation to full Bayesian inference, the FIM method has been widely used in GW astrophysics, particularly for the analysis of numerous simulated signals~\cite{Fisher1,Fisher2babak2016,klein2016science,Wangzm_2024,johnson2024source,Iacovelli2022mbg,population}. For representative high-confusion cases, we further apply Markov chain Monte Carlo (MCMC) method to perform full Bayesian parameter estimation~\cite{karnesis2023eryn,BayesianPEcost,PTMCMC}, thereby validating and extending the FIM-based predictions under more realistic conditions. 
The simulation of MBHB population indicates a relatively low rate of overlap, and severe confusion events are also found to be infrequent under different population models. 
Moreover, the inclusion of higher-order modes (HMs) significantly mitigates the impact of overlapping signals on parameter uncertainties, a result validated by MCMC analysis. In addition, the MCMC results further reveal that HMs play a key role in breaking parameter degeneracies, regardless of the presence of signal overlap. 
The findings presented here offer valuable insights for the development of data analysis strategies in future space-based gravitational-wave missions.

The remainder of this paper is organized as follows. 
In Sec.~\ref{sce:Prevalence}, we introduce the three representative MBHB population models and predict the prevalence of signal overlap, adopting the event rates justified earlier in this work.   Sec.~\ref{sec:Quantification} presents our methodologies and results for quantifying the impact of source confusion on parameter estimation. 
We begin in Sec.~\ref{sec:fisher_framework} with an overview of the FIM formalism. 
In Sec.~\ref{sec:fisher_reslut}, we apply this method to assess how overlapping signals affect parameter uncertainties, with particular emphasis on the role of HMs. Sec.~\ref{sec:MCMC_reslut} presents full Bayesian inference using MCMC for representative overlapping configurations, validating and extending the Fisher-based predictions. 
Finally, we summarize our main findings and discuss future implications in Sec.~\ref{sec:conclusion}. 

\section{The Prevalence of source confusion}
\label{sce:Prevalence}
\subsection{MBHB Population Models}
For MBHBs, the occurrence of source confusion is closely related to the distributions of their physical  parameters, especially their merger times and masses. 
MBHB populations might grow from rather distinct seeding mechanisms, each with unique mass distributions. 
To estimate the occurrence rate of overlapping MBHBs in the data stream, we consider three semi-analytical models~\cite{mangiagli2022massive}, including \textsc{Q3d},  \textsc{Q3nod} and \textsc{PopIII}~\cite{barausse2012evolution,klein2016science,population}, and obtain respectively the mass and spin distributions. 
These models consistently follow the mass and spin growth of massive black holes (MBHs) through accretion and feedback processes. 
In particular, \textsc{Q3d} and \textsc{Q3nod} models often predict high and partially aligned spins at the time of binary formation, though systems with low or misaligned spins can also occur~\cite{klein2016science}. 
The \textsc{\textsc{PopIII}} model assumes light MBH seeds formed from the remnants of Population III stars, incorporating delays between galaxy formation and MBH coalescence~\cite{POPIII}. The \textsc{Q3d} model considers heavy seeds originating from the collapse of protogalactic disks, with events limited to redshifts $z \lesssim 10$ and typical detector-frame chirp masses $\mathcal{M}_z \gtrsim 10^6\,M_\odot$~\cite{volonteri2008evolution}. 
The source-frame chirp mass is defined as  $\mathcal{M} \equiv (m_1 m_2)^{3/5}/(m_1 + m_2)^{1/5}$, where $m_1$ and $m_2$ are the component masses of the binary, and $\mathcal{M}_z \equiv (1 + z)\mathcal{M}$ with $z$ the redshift. 
The \textsc{Q3nod} model shares the same seeding mechanism as \textsc{Q3d} but ignores merger delays, \emph{i.e.} assumes that MBHs coalesce as soon as galaxies merge~\cite{Zhu_2024}, which results in a broader redshift distribution, reaching up to $z \sim 20$, and a wider mass range from $\mathcal{M}_z \sim 10^5\,M_\odot$ to $10^8\,M_\odot$~\cite{population,chen2020dynamical}, therefore a higher event rate.

\subsection{Overlapping Rate Prediction}
\label{sec:two_signals}
We begin by adopting the simulation framework developed in Ref.~\cite{johnson2024source}, which matches observed local merger rates with theoretical MBHB population models, while accounting for the delay between black hole formation and coalescence. This framework enables computation of both the total number of MBHB events per year ($N_{\mathrm{MBHB}}$) and the average time interval between successive events ($\langle \Delta t \rangle$, in days) across different assumptions for the local merger rate. Table~\ref{tab:rate_summary} presents the outcomes of our simulations , showing a strong dependence of event statistics on the assumed merger rate.

\begin{table}[h]
\begin{threeparttable}
\caption{Simulated MBHB event rates for a range of local merger rates, accounting for formation-to-merger delays.}
\label{tab:rate_summary}

\begin{tabular*}{\columnwidth}{@{\extracolsep{\fill}}ccc} 

\toprule
Rate [Gpc$^{-3}$\,yr$^{-1}$] & $N_{\mathrm{MBHB}}$ & $\Delta t$ [days]\\
\midrule
0.01 &   13 & 13.3 \\
0.05 &   71 & 2.5  \\
0.07 &  100 & 1.8  \\
0.5  &  723 & 0.3  \\
1.0  & 1444 & 0.1  \\
5.0  & 7238 & 0.03 \\
\bottomrule
\end{tabular*}
\end{threeparttable}
\end{table}

To characterize the prevalence of source confusions quantitatively across different MBHB populations, we conduct detailed numerical simulations based on the \textsc{PopIII}, \textsc{Q3d}, and \textsc{Q3nod} models. 
Before presenting the results, we briefly introduce several key aspects of our simulation methods. 
To model the GW signals, we follow the standard practice of decomposing the strain in the transverse-traceless gauge using spin-weighted spherical harmonics~\cite{2Ylm,Marsat51,Multipoleexpansions}: 
\begin{equation}
h_{+}-i h_{\times}=\sum_{\ell \geq 2} \sum_{m=-\ell}^{\ell}{ }_{-2} Y_{\ell m}(\iota, \varphi) h_{\ell m} .
\end{equation}
Here, $\iota$ and $\varphi$ denote the inclination and azimuthal angles in the source frame, respectively.  
Waveforms are generated using the \texttt{IMRPhenomHM} model implemented in the \texttt{BBHx} package~\cite{katz2020hku,Katz2022}, which includes HMs $(\ell, \vert m \vert) =  (3, 3),\ (4, 4),\ (2, 1),\ (3, 2),\ (4, 3)$, in addition to the dominant quadrupole mode  $(\ell = 2,\ \vert m \vert = 2)$. 
The \texttt{IMRPhenomD} waveform model, which includes only the dominant mode, is also used when HMs are neglected. 
These HMs become especially relevant for binaries with asymmetric masses or inclined orbital orientations~\cite{london2018first}.
 
One motivation for including HMs, which introduces additional structure to the waveform and thus improves parameter estimation and breaks degeneracies~\cite{yi2025systematicbiasesexclusionhigher,HMsLISA,HMs_slj,HMs_bbh,HMs_localization,Ng_2023}, is their potential to mitigate source confusion. 
As discussed previously, overlap between signals primarily depends on their time-frequency trajectories. Since each harmonic mode has a distinct frequency evolution,  and thereby helps to reduce confusion impacts. We investigate this hypothesis using both Fisher matrix analysis and full MCMC parameter inference
Additionally, given that the SNRs of MBHB signals can reach the order of $\mathcal{O}(10^3)$, it would be necessary to account for the  subdominant waveform features for unbiased estimation~\cite{Gong:2023ecg}.

All the generated signals are in the sensitive frequency band of Taiji, that from $10^{-4}$ to $0.5$ Hz. Table~\ref{tab:parameter_list} summarizes the waveform parameters used in the simulations. To simulate more realistic detector responses, we incorporate second-generation time-delay interferometry (TDI) channels, which apply light-travel delays to inter-spacecraft heterodyne interferometric phase measurements and linearly combine them to construct virtual equal-arm interferometers that effectively suppress laser frequency noise~\cite{Tinto:2004wu,PhysRevD.107.082001}. 
Throughout this study, we model the TDI response of MBHB signals following the formalism of Ref.~\cite{Marsat2021}, which is further adapted for Taiji's orbit and second-generation Michelson TDI-$X_2$ channel (see the Appendix of Ref.~\cite{Yuan:2025wyx} for details).

\begin{table}[h]
\begin{threeparttable}
\caption{List of source parameters used in the simulations, along with their corresponding symbols and physical units.}
\label{tab:parameter_list}

\begin{tabular*}{\columnwidth}{@{\extracolsep{\fill}}ccc} 
\toprule

Parameter           & Symbol                 & Unit  \\

\midrule

Detector-frame chirp mass& $\mathcal{M}_z$        & $M_\odot$  \\
Mass ratio                & $q$                    & -  \\
Primary spin (aligned)    & $\chi_{z1}$               & -  \\
Secondary spin (aligned)  & $\chi_{z2}$               & -  \\
Coalescence time          & $t_c$                  & day  \\
Coalescence phase         & $\Phi_c$               & rad  \\
Luminosity distance       & $D_L$                  & Mpc  \\
Inclination angle         & $\iota$                & rad  \\
Ecliptic longitude        & $\lambda$              & rad  \\
Ecliptic latitude         & $\beta$                & rad  \\
Polarization angle        & $\psi$                 & rad  \\

\bottomrule
\end{tabular*}
\end{threeparttable}
\end{table}

To quantify the prevalence of source confusion, the simulations assume a local merger rate of $0.07~\mathrm{Gpc}^{-3}\mathrm{yr}^{-1}$, corresponding to approximately 100 MBHB signals during a year observation. This lies within the commonly predicted range for space-based detectors. For instance, according to the mission definition report of LISA~\cite{LISA_redbook2024}, the expected detection rate of MBHB mergers, integrated over mass and redshift, ranges between a few and $\sim 100$ events per year, depending on the black hole seeding scenario and merger time delay. N. Steinle \emph{et al.}~\cite{ptaLISAdetectionsMBHB} constrain the 95\% upper limit of MBHB detections to be below 134  $\mathrm{yr}^{-1}$ for total masses within $10^7 - 10^8 M_\odot$, based on a Bayesian-inferred merger rate and an astrophysically motivated formation model.

In the simulation, the analysis window spans six months, with the coalescence time of the target signal placed at its center to avoid boundary effects due to signal truncation. This setup ensures that both the target signal and any potentially overlapping sources are fully captured, preventing artificial suppression or enhancement of confusion statistics.

\begin{figure}[!htbp]
\centering
\includegraphics[width=\columnwidth]{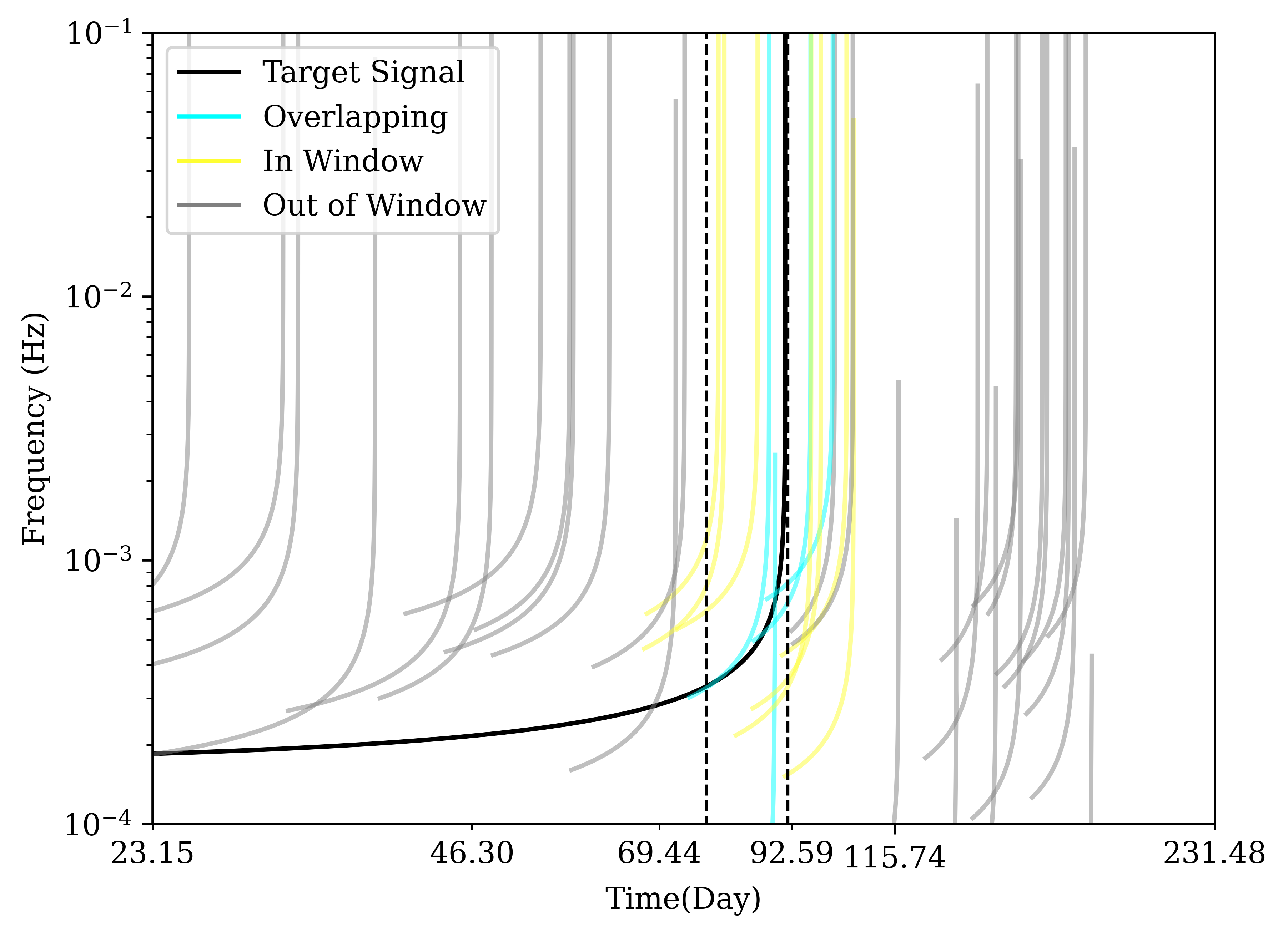}
\caption{Time-frequency map of a representative \textsc{PopIII} simulation. The non-target signals shown in the figure are limited to those with cumulative SNR exceeding 10. The black vertical dashed line marks the observation window, which covers a duration of 15 days and extends up to 0.5 days after the merger of the target signal. Yellow traces denote all signals within the observation window that coincide with the target signal (black) in the time domain, while blue traces indicate those that coincide with it simultaneously in both the time and frequency domains. Only the time–frequency trajectories of the dominant $(2,2)$ mode are shown, and overlap is evaluated based solely on their crossings.}
\label{figpopIII_example}
\end{figure}

Figure~\ref{figpopIII_example} provides a representative example from the \textsc{PopIII} population. Within the observation window, several signals (yellow) coincide with the target signal (black) in the time domain, only a small subset (blue) exhibits simultaneous overlap in both time and frequency domains, which we define as a ``genuine'' source confusion. 

To ensure that the mean overlap number of the simulations output is statistically stable, we estimate the required number of realizations using Cochran’s formula~\cite{Yaya2024SampleSize}, which assumes approximate normality of the sample mean as justified by the central limit theorem. The required number of simulations $N$ should be no less than 
\[
N \ge \left( \frac{Z \cdot {s}}{\epsilon} \right)^2,
\]
where $s$ is the estimated sample standard deviation, $\epsilon$ is the target margin of error, and $Z$ is the z-score corresponding to the desired confidence level. In our case, we set $\epsilon= 0.1$  for \textsc{PopIII} and $\epsilon= 0.05$ for both \textsc{Q3d} and \textsc{Q3nod}, and adopt $Z=2.33$ for a $98\%$ confidence level. 
For \textsc{PopIII}, 1285 realizations yield a mean overlap count of $4.184$  overlapping with target signal per year. For \textsc{Q3d}, 511 simulations yield an average of $0.314$ overlaps, and for \textsc{Q3nod}, 775 simulations result in an average of $0.348$ overlaps with the same statistical precision. These sample sizes are selected to ensure that the resulting estimates for the mean number of overlaps meet the targeted confidence bounds. Overall, source confusion among MBHBs appears to be relatively uncommon. Accordingly, the following analysis focuses exclusively on cases involving the overlap of two MBHB signals.

\section{Quantifying the Impact of Source Confusion }
\label{sec:Quantification}
We evaluate the impact of source confusion on MBHB parameter estimations using two complementary methods: the FIM analysis and MCMC sampling. The Fisher matrix approach provides rapid uncertainty estimates under the assumption of a high-SNR regime, where the posterior is approximately Gaussian.  

We define an uncertainty ratio $\gamma$ under the Fisher matrix formalism to quantify how source confusion affects the parameter estimation precision of a target signal. Based on the analytical model proposed in Ref.~\cite{johnson2024source}), we investigate how variations in the phase difference evaluated at the overlap point affect the uncertainties of parameters. We further compare results for different MBHB chirp-mass cases and assess how the inclusion or exclusion of HMs affects the degree of source confusion.

To assess the uncertainties introduced by source confusion under more realistic data analysis scenarios, we additionally perform full Bayesian inference using MCMC for selected configurations of auxiliary variables that lead to significantly elevated uncertainty ratios $\gamma$. These configurations are identified through Fisher matrix analysis as those in which source confusion has significant impact on parameter estimation precision.  

\subsection{The Fisher Formalism for Source Confusion}
\label{sec:fisher_framework}
The total data stream is modeled as the sum of $N$ GW signals and stationary, zero-mean Gaussian noise, which is a simplified assumption often used in theoretical analysis~\cite{2008GWbook}: 
\begin{equation}
    D(t) = \sum_{n=1}^{N} h_n(t; \theta_n) + \mathrm{Noise}(t),
\end{equation}
where $h_n(t; \theta_n)$ denotes the waveform of the $n$-th signal, and $\theta_n$ is its corresponding parameter set. The noise-weighted inner product~\cite{Flanagan_2005} is defined as 
\begin{equation}
    (h | g) \equiv  4 \operatorname{Re} \int_{0}^{\infty} \frac{\tilde{h}(f) \tilde{g}^*(f)}{S_n(f)} df.
\end{equation}
where tildes denote Fourier transforms and $S_n(f)$ is the one-sided power spectral density (PSD) of the detector noise.  The log-likelihood~\cite{Marsat2021} for a given parameter set $\bm{\Theta} = (\theta_1, \theta_2, ..., \theta_N)$ is given by: 
\begin{equation}
 \ln  \mathcal{L}  \propto  -\frac{1}{2} \left( D - H(\bm{\Theta}) \big| D - H(\bm{\Theta}) \right),
\end{equation}
with the total waveform defined as $ H(\bm{\Theta}) = \sum_{n=1}^{N} h_n(t; \theta_n) $.

Assuming a high-SNR case, the composite FIM~\cite{LISAsconfusion04} provides an efficient approximation for estimating parameter uncertainties: 
\begin{equation}\label{eq:Fisher_element}
    \Xi_{ij} = \left( \frac{\partial H}{\partial \Theta^i} \bigg| \frac{\partial H}{\partial \Theta^j} \right).
\end{equation}
and the corresponding covariance matrix, which determines the parameter uncertainties,  is given by 
\begin{equation}
    \bm{\Sigma} = \bm{\Xi}^{-1}.
\end{equation}
To evaluate how source confusion impacts parameter estimation, we consider a simplified scenario with two overlapping signals. The composite Fisher matrix in this case is structured as: 
\begin{equation}
    \bm{\Xi} =
    \begin{bmatrix}
        \bm{\xi}^{(1)} & \bm{C}^{(12)} \\
        \bm{C}^{(21)} & \bm{\xi}^{(2)}
    \end{bmatrix}.
\end{equation}
where $\bm{\xi}^{(1)}$ and $\bm{\xi}^{(2)}$ are the Fisher matrices for the individual signals, and 
\begin{equation}
    C_{ij}^{(12)} = \left( \frac{\partial h_1}{\partial \theta^i} \bigg| \frac{\partial h_2}{\partial \theta^j} \right)
\end{equation}
represents the cross-term capturing signal interference.

We define the uncertainty ratio $\gamma$ to quantify the impact on the uncertainties of the parameters due to source confusion. Specifically, we first compute the uncertainty $\Delta\theta^i$ for an isolated target signal. The second signal is then introduced to partially overlap the target, and the corresponding uncertainty under confusion is denoted $\Delta\Theta^i$. 
The ratio 
\begin{equation}
    \gamma^i = \frac{\Delta\Theta^i}{\Delta\theta^i} = \sqrt{\frac{(\bm{\Xi}^{-1})^{ii}}{(\bm{\xi}^{-1})^{ii}}},
\end{equation}
is computed from the diagonal elements of the inverse Fisher matrices.
This formulation provides a theoretical framework for assessing the influence of signal overlap on parameter uncertainties. 

\subsection{Source Confusion Analysis with Fisher Matrix}

Previous studies have demonstrated that parameter uncertainties can increase substantially when two GW signals exhibit similar time-frequency evolution, a feature largely dictated by their detector-frame chirp masses \cite{Wangzm_2024,crossterm06,johnson2024source,sourceR21}. Signals with comparable values of $\mathcal{M}_z$ tend to overlap in broad frequency bands, enhancing cross-correlations and thus increasing the probability of confusion. To assess such impacts, we begin by exploring a range of chirp mass from $10^4$ to $10^7~M_\odot$ that are relevant to space-based detectors.

Moreover, to assess how source confusion affects parameter estimation, we follow Ref.~\cite{johnson2024source}, which examined how the resulting parameter uncertainties vary with three key quantities associated with overlapping signals: 
(1) the overlap frequency $f_{\mathrm{ov}} $,
(2) the phase difference at overlap, and $\Delta\Phi(f_{\mathrm{ov}})=\Phi_{1}-\Phi_{2}$, 
(3) the chirp mass difference, $\Delta\mathcal{M}_z = |\mathcal{M}_{z1} - \mathcal{M}_{z2}|$. 

In this study, we focus on $\Delta\mathcal{M}_z $ and $\Delta\Phi(f_{\mathrm{ov}})$. 
The former determines the degree of similarity between two signals, whereas the phase difference itself plays a particularly critical role, since gravitational-wave parameter inference, especially for intrinsic parameters, relies primarily on phase matching~\cite{phaseevolution_contains_more_information,Phase_importance}. 
Hence, the phase difference at the overlap frequency is expected to be a decisive factor in determining the degradation of parameter precision when signals overlap in the time–frequency domain. 
To characterize these dependencies, we evaluate the parameter-uncertainty ratio for the target signal by independently varying $\Delta\mathcal{M}_z$ and $\Delta\Phi$, with  $f_{\mathrm{ov}}$ controlled (following the rule given in Sec.~\ref{subsec:Mz_magnitude})  throughout the analysis.

For simplicity, we focus on a representative \textbf{Target Signal} from a typical coalescing MBHB detectable by the Taiji mission, and the other \textbf{Overlapping Signal} is then constructed with 
parameters that closely match those of the target. 
This setup serves as a baseline for subsequent Fisher matrix studies, in which variations will be systematically made to explore the impact of source confusion on parameter estimation. The full parameter sets are listed below:
\begin{itemize}
    \item \textbf{Target Signal} parameters: 
    $\mathcal{M}_z = 10^4\sim 10^7$, $q = 0.87$, $\chi_{z1} = 0.90$, $\chi_{z2} = 0.92$, $t_c = 30.0$, $\Phi_c = 0.50$, $D_L = 30{,}200$, $\iota = 0.88$, $\lambda = 3.35$, $\beta = 0.12$, $\psi = 1.56$;
    \item \textbf{Overlapping Signal} parameters: 
    $\mathcal{M}_z = 10^4\sim 10^7$, $q = 0.87$, $\chi_{z1} = 0.899$, $\chi_{z2} = 0.921$, $t_c = 30.0$, $\Phi_c = 0.58$, $D_L = 30{,}210$, $\iota = 0.60$, $\lambda = 3.338$, $\beta = 0.11$, $\psi = 1.68$.
\end{itemize}

Unless otherwise specified, the \textbf{Target signal} parameters are kept fixed and only the $\mathcal{M}_z$, $t_c$ and $\Phi_c$ are varied. Adjusting the coalescence time of the \textbf{Overlapping Signal} shifts the time-frequency trajectory along the time axis, thereby changing $f_{\mathrm{ov}}$. 
Modifying the coalescence phase of \textbf{Overlapping Signal} directly alters the phase difference at overlap $\Delta\Phi(f_{\mathrm{ov}})$. 

We use the \texttt{fishertools} module from the \texttt{gwfast} package to invert the Fisher matrix~\cite{Iacovelli2022bbs,Iacovelli2022mbg}, thereby obtaining the covariance matrix. The subsequent subsections present investigations of various influencing factors, with particular emphasis on the impact of HMs. 
As summarized in Ref.~\cite{cornish2025tdifly}, 
any gravitational-wave strain can be represented in an abstract form: 
\[
h(t) = \Re\!\left[\sum_{n} A_n(t)\,e^{i\Phi_n(t)}\right],
\]
where \(A_n(t)\) are the real amplitude tensors for each harmonic mode and \(\Phi_n(t)\) are the corresponding phases. 
For each mode, the amplitude and phase evolve smoothly and slowly over time.  
Since different modes exhibit distinct time-frequency evolutions, our following analysis will use the dominant (2,2) mode to determine $f_{\mathrm{ov}}$ and $\Delta\Psi(f_{\mathrm{ov}})$. 

\subsubsection{Impact of Chirp Mass Magnitude}\label{subsec:Mz_magnitude}

\label{sec:fisher_reslut}
\begin{figure*}[t]
\centering
\includegraphics[width=0.75\textwidth]{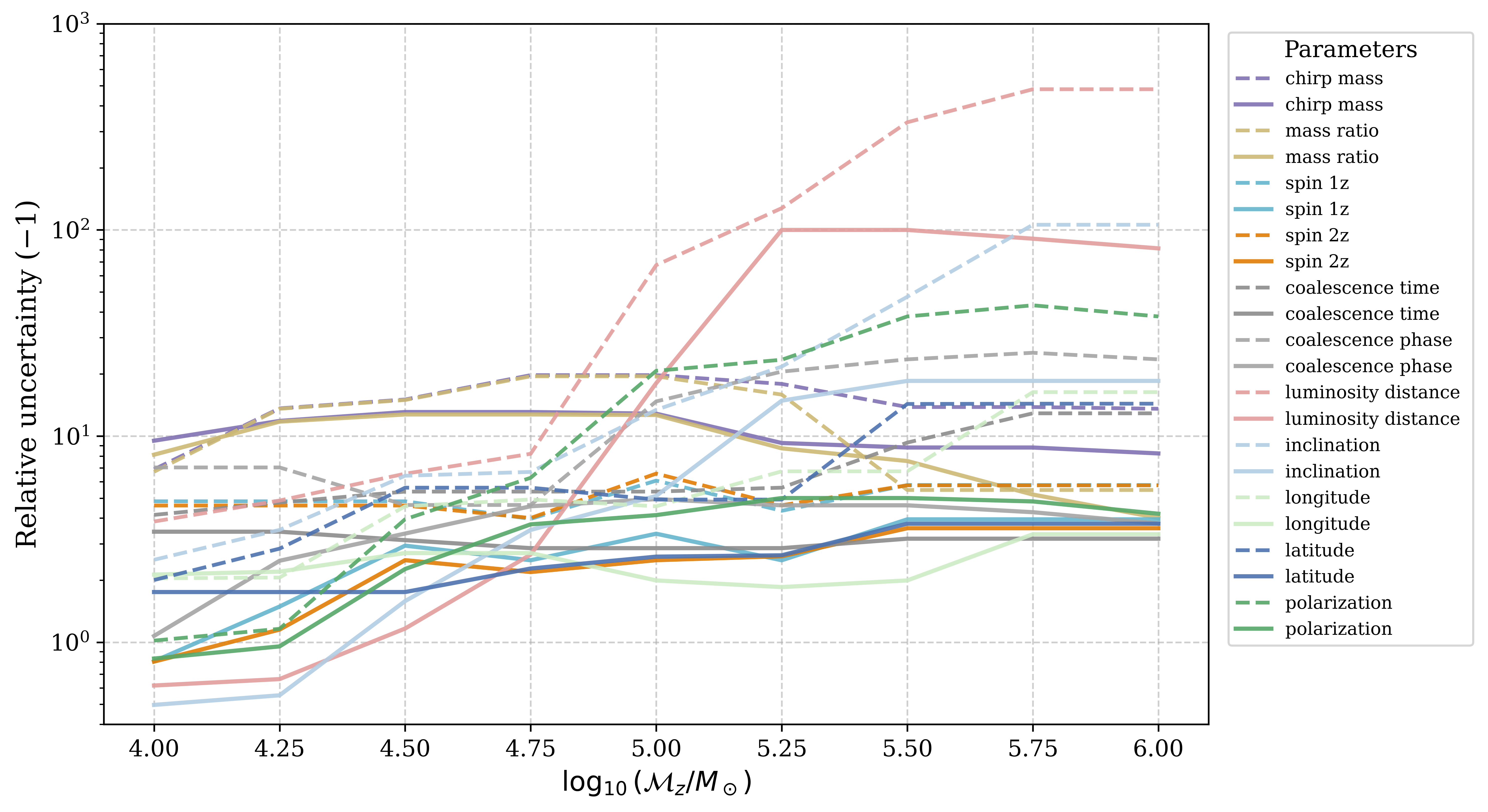}
\caption{Parameter uncertainty ratio $\gamma^i - 1$ as a function of chirp mass scale for the  case $\Delta\mathcal{M}_z = 0$. Solid lines represent waveforms including HMs; dashed lines show results for the dominant $(2,2)$ mode only. The phase difference at the point of overlap is fixed ($\Delta\Phi (f_{\mathrm{ov}}) = 0$), and $f_{\mathrm{ov}}$ is chosen to scale inversely with the $\mathcal{M}_z $.}
\label{fig:massscale_gamma}
\end{figure*}

Figure~\ref{fig:massscale_gamma} presents the parameter uncertainty ratio $\gamma^i-1$ under the condition of $\Delta\mathcal{M}_z = 0$, evaluated on a range of chirp mass scales ($10^4\sim 10^6~M_\odot$). 
Dashed curves correspond to the $(2,2)$ mode waveform, while the solid curves include HMs. Since the SNR mainly accumulates near coalescence, we adopt a fixed observation duration (from five days before to half a day after merger). 
MBHBs with larger $\mathcal{M}_z$ evolve more rapidly in frequency as the frequency evolves 
\[
df/dt =   \frac{96}{5} \pi^{8/3}\mathcal{M}_z^{5/3} f^{11/3}
\]
at the leading order~\cite{Cutler_1994}. 
Their inspiral therefore sweeps through the detector band more quickly and merges at lower frequencies. The frequency at a fixed time before coalescence scales as as $f\propto \mathcal{M}_z^{-5/8}$, accordingly, we impose the same mass dependence on the overlap frequency $f_{\mathrm{ov}}$ to keep the overlap impact comparable across different masses. 
It is worth noting that when $\Delta\mathcal{M}_z = 0$ and $\mathcal{M}_z$ exceeds $10^6~M_\odot$, the corresponding condition numbers of FIM are also extremely high, indicating that the Fisher estimates in this region are not reliable. 

The resulting uncertainty ratio $\gamma^i-1$ generally increases with the chirp mass and can reach magnitudes of $\sim10^2$ in the high–mass regime when only the $(2,2)$ mode is considered.
When HMs are included, each mode carries a distinct frequency evolutions, and there is no unique overlap frequency or overlap phase, leading to slight fluctuations in the results. 
Even in cases where two signals are nearly identical in all intrinsic parameters and thus exhibit highly overlapping time-frequency structures, the uncertainty increase remains within a factor of $\mathcal{O}(10^1)$ for most parameters relative to the isolated-signal case. These results highlight the importance of HMs in mitigating the adverse effects of source confusion.

\subsubsection{Impact of Chirp Mass and Phase Differences}

\begin{figure*}
\centering
\includegraphics[width=0.95\textwidth]{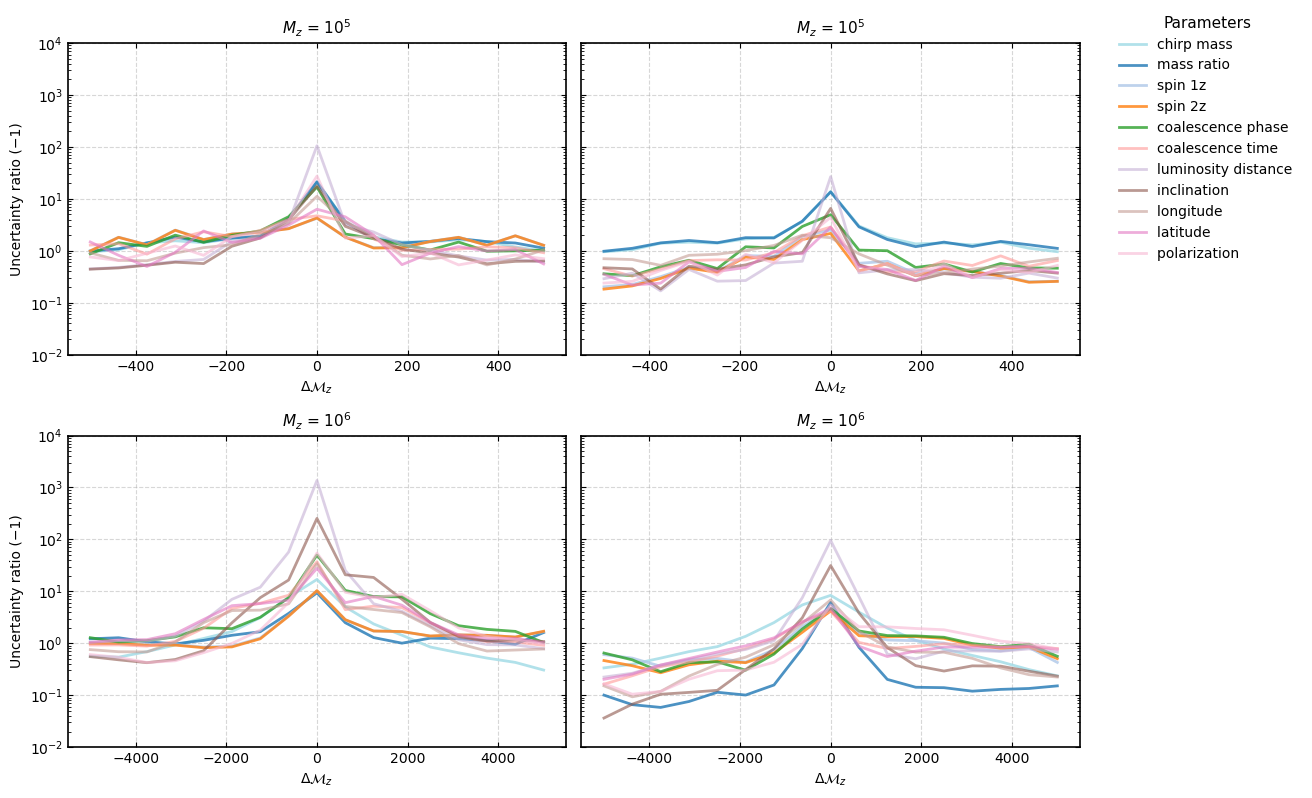}
\caption{Parameter uncertainty ratio $\gamma^i - 1$ as a function of $\Delta\mathcal{M}_z$ for the baseline parameter sets with $\mathcal{M}_z = 10^5~M_\odot$ (top) and $\mathcal{M}_z = 10^6~M_\odot$ (bottom). Left panels show results using only the dominant $(2,2)$ mode, while right panels include HMs. The variation range of $\Delta\mathcal{M}_z$ corresponds to about 1\% of each magnitude of chirp mass. All cases assume $\Delta\Phi (f_{\mathrm{ov}})= 0$}
\label{fig:deltaMz_gamma}
\end{figure*}

We analyze the variation of the parameter uncertainty ratio $\gamma^i-1$ with respect to the chirp mass difference $\Delta \mathcal{M}_z$, focusing on two representative scales: $\mathcal{M}_z = 10^5$ and $10^6~M_\odot$ (Fig.~\ref{fig:deltaMz_gamma}). The maximum uncertainty ratio $\gamma^i-1$ appears at $\Delta\mathcal{M}_z = 0$, indicating that parameter confusion peaks when chirp masses are identical~\cite{sourceR21,johnson2024source}. Incorporating HMs mitigates this effect, particularly at higher $\mathcal{M}_z$. 
With HMs included, the uncertainty ratio $\gamma^i - 1$ can exceeds $\mathcal{O}(10^1)$ when $\Delta \mathcal{M}_z / \mathcal{M}_z$ falls below 0.2\%, implying that source confusion enhance parameter uncertainties but typically within one order of magnitude compared to the non-confused case. For chirp mass scale $10^5~M_\odot$, the uncertainty ratio $\gamma^i - 1$ is remains mild, with $\gamma^i - 1$ below $\mathcal{O}(10^0)$ for most parameters when $\Delta \mathcal{M}_z / \mathcal{M}_z < 0.2\%$. 

Based on numerical simulations of the population models \textsc{PopIII}, \textsc{Q3d}, and \textsc{Q3nod}, after excluding binaries outside the Taiji sensitivity band, we find that the fraction of two MBHB signals with $\Delta \mathcal{M}_z / \mathcal{M}_z<0.2\%$ (relative to the smaller chirp mass in the pair) is 0.14\%, 0. 08\% and 0.12\%, respectively. 
These results indicate that, even under the assumption of near-identical signal parameters, such extreme mass degeneracies are rare in these three MBHB populations. 
Consequently, severe source overlap can generally be neglected in MBHB data analysis.

\begin{figure}[b]

\centering
\includegraphics[width=\columnwidth]{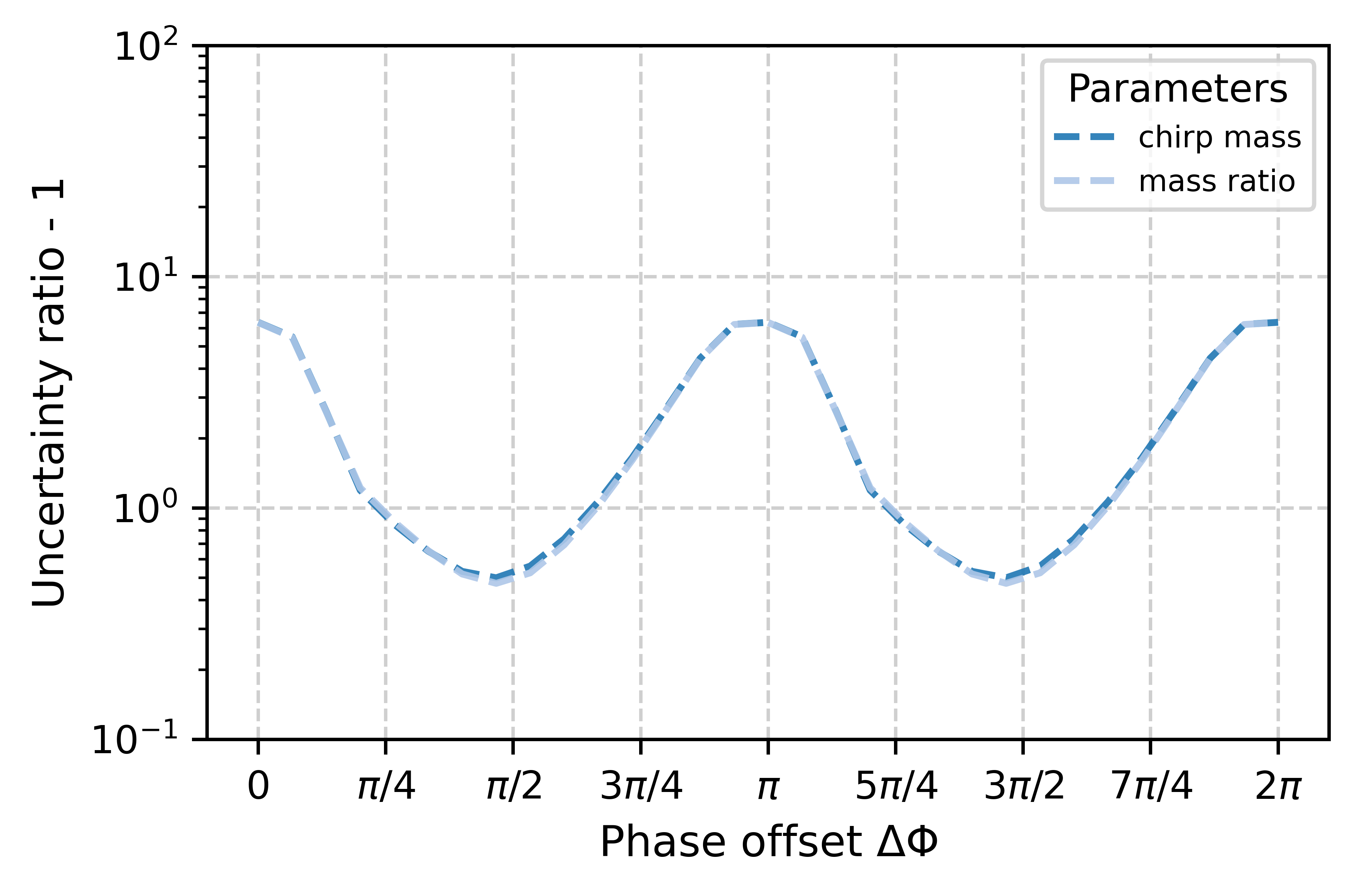}

\caption{Parameter uncertainty ratio $\gamma^i - 1$ as a function of phase difference $\Delta\Phi=\Phi_1 - \Phi_2 $ for the baseline parameter sets with $\mathcal{M}_z = 10^5\,M_\odot$. Dashed lines represent waveforms with the $(2,2)$ mode only. $\Phi_1$ and $\Phi_2$ are the signal phases in the TDI-$X_2$ channel. The $f_{\mathrm{ov}}$ is fixed and $\Delta \mathcal{M}_z = 0$.}
\label{fig:phase_gamma}
\end{figure}

Figure~\ref{fig:phase_gamma} shows the variation of $\gamma^i - 1$ as a function of the phase difference $\Delta\Phi= \Phi_1 - \Phi_2$ at $f_{\mathrm{ov}}$ when $\mathcal{M}_z = 10^5~M_\odot$. Here, only the two parameters most sensitive to phase evolution, the chirp mass and the mass ratio, are shown. A nearly symmetric pattern is observed between the intervals $[0, \pi]$ and $[\pi, 2\pi]$ for most parameters, with $\gamma^i - 1$ reaching a local maximum near $\Delta\Phi = 0$, decreasing toward a minimum around $\Delta\Phi = \pi/2$, and rising again thereafter. This behavior arises because for the dominant $(2,2)$ mode, phase differences of $\Delta\Phi = 0$ or $\pi$ align the two waveforms, maximizing their similarity and thus enhancing parameter degeneracy, which leads to the largest uncertainties. Even in the extreme case of $\Delta \mathcal{M}_z = 0$, the results show that the inclusion of HMs keeps $\gamma^i - 1$ within $\mathcal{O}(10^1)$ for these two parameters. 

\subsection{Source Confusion Analysis with Full Bayesian Inference}
\label{sec:MCMC_reslut}
To explore the influence of source confusion on parameter estimations in more realistic data analysis scenarios, based on the theoretical uncertainties obtained via Fisher matrix, we select a representative parameter setting for two signals having significant source confusion. Specifically, we consider the setup with $\mathcal{M}_z = 10^6$ and $\Delta\Phi = 0$, where the uncertainty ratio $\gamma^i-1$ reaches relatively high values. 
Parameters of the \textbf{Target Signal} and \textbf{Overlap Signal} can be found in the second column of Table~\ref{tab:mcmc_fisher_22d}. 

\begin{table*}[htb]
\centering 
\begin{threeparttable}
\caption{Posterior modes (68\% credible intervals) from MCMC sampling and $1 \sigma$ uncertainties from the Fisher matrix for 11 parameters, comparing waveforms with and without HMs. Injected (true) values are shown for reference. The parameters of the \textbf{Target Signal} are the same as the previous settings, except that $\mathcal{M}_z$. }
\label{tab:comparison_mcmc/Fisher11}

\begin{tabular*}{\textwidth}{@{\extracolsep{\fill}}cccccc}
\toprule
\hline

        \multirow{2}{*}{Parameter}
         & \multirow{2}{*}{Injected value} 
         & \multicolumn{2}{c}{MCMC posterior (68\% credible interval)}
         & \multicolumn{2}{c}{Fisher 1$\sigma$ uncertainty} \\
         &  & {Without HMs} & {With HMs} & {Without HMs} & {With HMs} \\

\midrule

$\lg M_z$ & $6.00$ &
$6.000000^{+1.71\times10^{-5}}_{-1.72\times10^{-5}}$ &
$6.000000^{+1.57\times10^{-5}}_{-1.56\times10^{-5}}$ &
$\pm 1.70\times10^{-5}$ & $\pm 1.55\times10^{-5}$ \\

$q$ & $0.87$ &
$0.870644^{+5.52\times10^{-3}}_{-5.20\times10^{-3}}$ &
$0.869999^{+5.69\times10^{-4}}_{-5.68\times10^{-4}}$ &
$\pm 4.65\times10^{-3}$ & $\pm 5.75\times10^{-4}$ \\

$\chi_{z1}$ & $0.90$ &
$0.898619^{+1.01\times10^{-2}}_{-1.11\times10^{-2}}$ &
$0.899987^{+6.96\times10^{-4}}_{-6.89\times10^{-4}}$ &
$\pm 8.85\times10^{-3}$ & $\pm 6.94\times10^{-4}$ \\

$\chi_{z2}$ & $0.92$ &
$0.921548^{+1.31\times10^{-2}}_{-1.22\times10^{-2}}$ &
$0.920012^{+1.02\times10^{-3}}_{-1.03\times10^{-3}}$ &
$\pm 1.07\times10^{-2}$ & $\pm 1.03\times10^{-3}$ \\

$t_c$ & $30.0$ &
$30.000000^{+6.55\times10^{-6}}_{-6.65\times10^{-6}}$ &
$30.000000^{+4.29\times10^{-6}}_{-4.30\times10^{-6}}$ &
$\pm 4.92\times10^{-6}$ & $\pm 3.90\times10^{-6}$ \\

$\Phi_c$ & $0.50$ &
$0.498746^{+1.16\times10^{-2}}_{-1.21\times10^{-2}}$ &
$0.500069^{+2.46\times10^{-3}}_{-2.50\times10^{-3}}$ &
$\pm 1.09\times10^{-2}$ & $\pm 2.45\times10^{-3}$ \\

$\lg D_L$ & $4.48000694$ &
$4.48003^{+1.86\times10^{-3}}_{-1.89\times10^{-3}}$ &
$4.48003^{+1.51\times10^{-3}}_{-1.47\times10^{-3}}$ &
$\pm 1.76\times10^{-3}$ & $\pm 1.30\times10^{-3}$ \\

$\cos(\iota)$ & $0.63715114$ &
$0.637279^{+1.86\times10^{-3}}_{-1.91\times10^{-3}}$ &
$0.637177^{+9.31\times10^{-4}}_{-9.44\times10^{-4}}$ &
$\pm 1.88\times10^{-3}$ & $\pm 9.38\times10^{-4}$ \\

$\lambda$ & $3.35$ &
$3.34942^{+7.52\times10^{-3}}_{-7.42\times10^{-3}}$ &
$3.34989^{+4.50\times10^{-3}}_{-4.69\times10^{-3}}$ &
$\pm 7.58\times10^{-3}$ & $\pm 4.62\times10^{-3}$ \\

$\sin(\beta)$ & $0.11971221$ &
$0.119834^{+6.38\times10^{-3}}_{-6.41\times10^{-3}}$ &
$0.119683^{+4.98\times10^{-3}}_{-5.16\times10^{-3}}$ &
$\pm 6.13\times10^{-3}$ & $\pm 4.61\times10^{-3}$ \\

$\psi$ & $1.56$ &
$1.56012^{+4.39\times10^{-3}}_{-4.42\times10^{-3}}$ &
$1.56009^{+3.30\times10^{-3}}_{-3.27\times10^{-3}}$ &
$\pm 4.41\times10^{-3}$ & $\pm 3.13\times10^{-3}$ \\
\hline
        \bottomrule
    \end{tabular*}
    \end{threeparttable}
\end{table*}

\subsubsection{Parameter Estimation and Analysis of Single-Signal}

\begin{figure*}[t]
\centering
\includegraphics[width=0.90\textwidth]{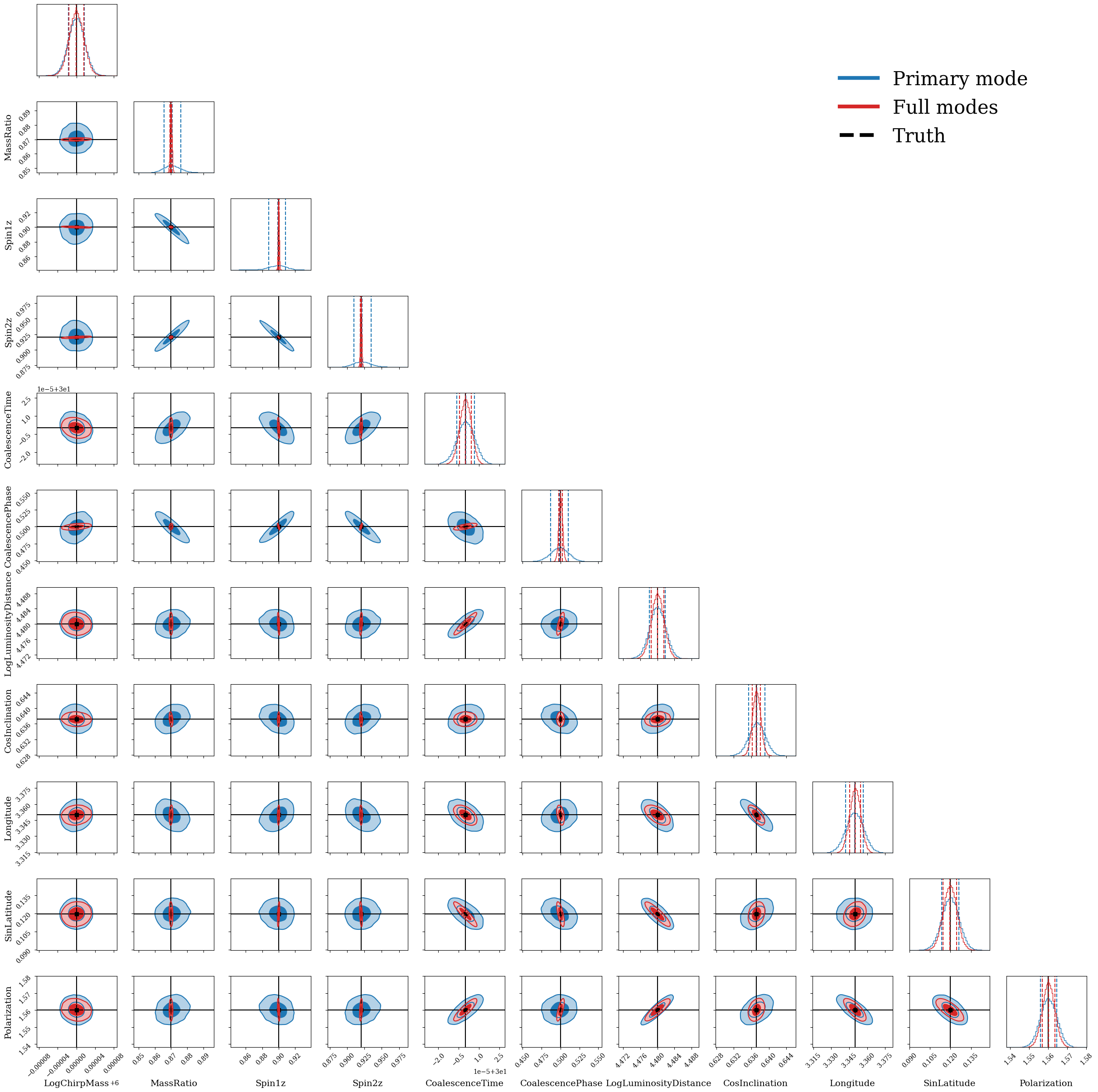}
\caption{Corner plot of the posterior distribution for a single MBHB signal using MCMC sampling. Red contours indicate results with HMs; blue contours correspond to the dominant $(2,2)$ mode only. The shaded regions represent the 1$\sigma$ and 2$\sigma$ credible intervals.}
\label{fig:single_corner}
\end{figure*}

We begin by analyzing a single signal using full MCMC sampling to estimate the posterior distributions of the source parameters. The injected values are listed in Table~\ref{tab:comparison_mcmc/Fisher11}, along with the numerical results for all 11 parameters. For each parameter, we report the posterior mode and 68\% credible interval derived from the MCMC sampling, as well as the corresponding $1\sigma$ uncertainty estimated using the Fisher matrix. This comparison not only offers a concise summary of both inference methods, but also provides valuable insight into the role of HMs in sharpening parameter uncertainties. The MCMC results are consistent with those obtained from the Fisher formalism in this single-source case, indicating that the Fisher matrix remains a reliable tool for uncertainty estimation when analyzing isolated MBHB signals.

For a more intuitive demonstration of the improvements brought from the HMs, Figure~\ref{fig:single_corner} shows the corner plot of the \textbf{Target signal}, with contours indicating the credible intervals of 1$\sigma$ and 2$\sigma$. The red curves correspond to waveforms including HMs, while the blue curves use only the dominant $(2,2)$ mode. 
Including HMs significantly reduces overall uncertainties in the estimated source parameters and effectively breaks degeneracies among intrinsic quantities, such as the mass ratio and component spins. In particular, HMs help to disentangle the correlation between luminosity distance and inclination angle, while also improving the precision of sky localization. These improvements are beneficial for facilitating multi-messenger follow-up observations.

\subsubsection{Parameter Estimation and Analysis of Overlapping-Signals}
We next explore a more complex situation involving two overlapping MBHB signals. Full Bayesian inference is performed using two independent waveform templates simultaneously matched to the combined data stream. The resulting posterior modes and 68\% credible intervals for each parameter are presented in Table~\ref{tab:mcmc_fisher_22d}, alongside the corresponding $1\sigma$ uncertainties derived from the Fisher matrix, allowing for direct comparison between the two inference methods.

\begin{table*}[htb]
\centering 
\begin{threeparttable}
\caption{MCMC posterior modes with 68\% central credible intervals and Fisher $1\sigma$ uncertainties for the two‐signal parameter estimation. Parameters without subscript denote the \textbf{Target signal}; parameters with subscript \emph{ov} denote the \textbf{Overlapping Signal}.}\label{table:injection}
\label{tab:mcmc_fisher_22d}

\begin{tabular*}{\textwidth}{@{\extracolsep{\fill}}cccccc}
\toprule
\hline

        \multirow{2}{*}{Parameter}
         & \multirow{2}{*}{Injected value} 
         & \multicolumn{2}{c}{MCMC posterior (68\% credible interval)}
         & \multicolumn{2}{c}{Fisher 1$\sigma$ uncertainty} \\
         &  & {Without HMs} & {With HMs} & {Without HMs} & {With HMs} \\

\midrule

$\lg M_z$ & 6.00  
  & $5.99992^{+1.46\times10^{-4}}_{-1.27\times10^{-4}}$  
  & $5.99994^{+1.13\times10^{-4}}_{-1.20\times10^{-4}}$  
  & $\pm\,4.80\times10^{-4}$ & $\pm\,2.14\times10^{-4}$ \\

$q$ & 0.87  
  & $0.870422^{+6.12\times10^{-3}}_{-5.57\times10^{-3}}$  
  & $0.867817^{+3.57\times10^{-3}}_{-3.58\times10^{-3}}$  
  & $\pm\,1.94\times10^{-1}$ & $\pm\,5.19\times10^{-3}$ \\

$\chi_{z1}$ & 0.90  
  & $0.898769^{+8.72\times10^{-3}}_{-1.00\times10^{-2}}$  
  & $0.902691^{+3.75\times10^{-3}}_{-4.32\times10^{-3}}$  
  & $\pm\,4.84\times10^{-1}$ & $\pm\,6.24\times10^{-3}$ \\

$\chi_{z2}$ & 0.92  
  & $0.919787^{+1.20\times10^{-2}}_{-9.44\times10^{-3}}$  
  & $0.915655^{+7.02\times10^{-3}}_{-6.47\times10^{-3}}$  
  & $\pm\,5.66\times10^{-1}$ & $\pm\,1.02\times10^{-2}$ \\

$t_c$ & 30.0  
  & $30.000000^{+1.48\times10^{-5}}_{-1.20\times10^{-5}}$  
  & $30.000000^{+1.22\times10^{-5}}_{-1.40\times10^{-5}}$  
  & $\pm\,6.40\times10^{-4}$ & $\pm\,3.27\times10^{-5}$ \\

$\Phi_c$ & 0.50  
  & $0.481023^{+3.97\times10^{-2}}_{-3.46\times10^{-2}}$  
  & $0.493661^{+1.08\times10^{-2}}_{-1.15\times10^{-2}}$  
  & $\pm\,4.99\times10^{0}$ & $\pm\,3.13\times10^{-2}$ \\

$\lg D_L$ & 4.48000694  
  & $4.47328^{+1.62\times10^{-2}}_{-1.96\times10^{-2}}$  
  & $4.46577^{+1.99\times10^{-2}}_{-1.63\times10^{-2}}$  
  & $\pm\,9.51\times10^{1}$ & $\pm\,2.11\times10^{-1}$ \\

$\cos(\iota)$ & 0.63715114  
  & $0.632794^{+1.45\times10^{-2}}_{-1.10\times10^{-2}}$  
  & $0.639469^{+4.85\times10^{-3}}_{-5.19\times10^{-3}}$  
  & $\pm\,1.55\times10^{1}$ & $\pm\,5.17\times10^{-2}$ \\

$\lambda$ & 3.35  
  & $3.34227^{+3.10\times10^{-2}}_{-3.05\times10^{-2}}$  
  & $3.35368^{+1.76\times10^{-2}}_{-2.20\times10^{-2}}$  
  & $\pm\,2.23\times10^{0}$ & $\pm\,5.54\times10^{-2}$ \\

$\sin(\beta)$ & 0.11971221  
  & $0.118071^{+1.46\times10^{-2}}_{-1.59\times10^{-2}}$  
  & $0.122705^{+1.39\times10^{-2}}_{-1.47\times10^{-2}}$  
  & $\pm\,7.55\times10^{-1}$ & $\pm\,4.06\times10^{-2}$ \\

$\psi$ & 1.56  
  & $1.55044^{+2.11\times10^{-2}}_{-2.93\times10^{-2}}$  
  & $1.55620^{+6.94\times10^{-3}}_{-6.78\times10^{-3}}$  
  & $\pm\,2.34\times10^{0}$ & $\pm\,2.70\times10^{-2}$ \\

$\lg M_{c,\mathrm{ov}}$ & 6.00  
  & $6.00006^{+1.14\times10^{-4}}_{-1.16\times10^{-4}}$  
  & $6.00005^{+9.95\times10^{-5}}_{-1.01\times10^{-4}}$  
  & $\pm\,3.86\times10^{-4}$ & $\pm\,1.75\times10^{-4}$ \\

$q_{\mathrm{ov}}$ & 0.87  
  & $0.869761^{+5.33\times10^{-3}}_{-4.56\times10^{-3}}$  
  & $0.872632^{+4.02\times10^{-3}}_{-4.01\times10^{-3}}$  
  & $\pm\,1.54\times10^{-1}$ & $\pm\,5.16\times10^{-3}$ \\

$\chi_{z1,\mathrm{ov}}$ & 0.899  
  & $0.900198^{+8.12\times10^{-3}}_{-7.98\times10^{-3}}$  
  & $0.894829^{+4.45\times10^{-3}}_{-3.96\times10^{-3}}$  
  & $\pm\,3.09\times10^{-1}$ & $\pm\,6.84\times10^{-3}$ \\

$\chi_{z2,\mathrm{ov}}$ & 0.921  
  & $0.920391^{+8.34\times10^{-3}}_{-8.11\times10^{-3}}$  
  & $0.927111^{+5.91\times10^{-3}}_{-6.79\times10^{-3}}$  
  & $\pm\,3.73\times10^{-1}$ & $\pm\,1.02\times10^{-2}$ \\

$t_{c,\mathrm{ov}}$ & 29.99999855  
  & $30.00000^{+1.06\times10^{-5}}_{-1.37\times10^{-5}}$  
  & $30.00000^{+9.15\times10^{-6}}_{-9.35\times10^{-6}}$  
  & $\pm\,3.92\times10^{-4}$ & $\pm\,2.37\times10^{-5}$ \\

$\Phi_{c,\mathrm{ov}}$ & 0.644441789  
  & $0.714801^{+7.29\times10^{-2}}_{-3.98\times10^{-2}}$  
  & $0.660353^{+1.54\times10^{-2}}_{-8.12\times10^{-3}}$  
  & $\pm\,6.03\times10^{1}$ & $\pm\,5.83\times10^{-2}$ \\

$\lg D_{L,\mathrm{ov}}$ & 4.48015073  
  & $4.48515^{+1.89\times10^{-2}}_{-2.44\times10^{-2}}$  
  & $4.49399^{+1.75\times10^{-2}}_{-2.00\times10^{-2}}$  
  & $\pm\,9.54\times10^{1}$ & $\pm\,2.12\times10^{-1}$ \\

$\cos(\iota_{\mathrm{ov}})$ & 0.82533561  
  & $0.827449^{+3.01\times10^{-2}}_{-3.51\times10^{-2}}$  
  & $0.827927^{+4.10\times10^{-3}}_{-4.42\times10^{-3}}$  
  & $\pm\,2.69\times10^{1}$ & $\pm\,4.31\times10^{-2}$ \\

$\lambda_{\mathrm{ov}}$ & 3.338  
  & $3.34019^{+2.75\times10^{-2}}_{-2.93\times10^{-2}}$  
  & $3.33371^{+1.78\times10^{-2}}_{-1.71\times10^{-2}}$  
  & $\pm\,8.25\times10^{-1}$ & $\pm\,4.01\times10^{-2}$ \\

$\sin(\beta_{\mathrm{ov}})$ & 0.1097783  
  & $0.116099^{+1.28\times10^{-2}}_{-1.60\times10^{-2}}$  
  & $0.107628^{+1.24\times10^{-2}}_{-1.11\times10^{-2}}$  
  & $\pm\,1.68\times10^{0}$ & $\pm\,3.13\times10^{-2}$ \\

$\psi_{\mathrm{ov}}$ & 1.68  
  & $1.7587^{+6.19\times10^{-2}}_{-5.88\times10^{-2}}$  
  & $1.69685^{+1.57\times10^{-2}}_{-1.37\times10^{-2}}$  
  & $\pm\,5.78\times10^{1}$ & $\pm\,5.15\times10^{-2}$ \\

\hline
        \bottomrule
    \end{tabular*}
    \end{threeparttable}
\end{table*}

By comparing the MCMC results of the \textbf{Target signal} in Tables~\ref{tab:comparison_mcmc/Fisher11} and Table~\ref{tab:mcmc_fisher_22d}, we find that the presence of an overlapping source typically amplifies the uncertainties of most parameters by factors ranges from a few to approximately ten, and rarely exceeding a factor of $\sim15$.
For example, including HMs, the uncertainty in the chirp mass ($\lg M_c$) grows from approximately $\sim1.6\times10^{-5}$ to about $\sim1.1\times10^{-4}$, and the uncertainty in the mass ratio ($q$) increases from around $\sim5.7\times10^{-4}$ to about $\sim3.6\times10^{-3}$. This confirms earlier FIM-based predictions and suggests that, even under relatively strong source confusion, the resulting parameter uncertainties, while larger, remain moderate with the help of HMs.

Moreover, for this specific parameter set, the comparison between MCMC and Fisher results  indicates that neglecting HMs leads to the Fisher matrix to overestimate uncertainties.
When HMs are excluded, the Fisher-matrix uncertainties of parameters such as $q$, $\chi_{z1}$, $\chi_{z2}$, and sky-location largely exceed those obtained from MCMC.
This discrepancy arises because, as mentioned before, we have selected events with severe overlap to perform the full Bayesian analysis. 
For this extreme case, 
the parameter degeneracy caused by source confusion leads to ill-conditioned Fisher matrix in the absence of HMs. 
Conversely, when HMs are included, Fisher-based uncertainties closely align with MCMC results. These findings highlight the crucial role of HMs in enabling accurate Fisher-matrix inferences, particularly under significant signal overlap conditions. 

\begin{figure*}[t]
\centering
\includegraphics[width=0.90\textwidth]{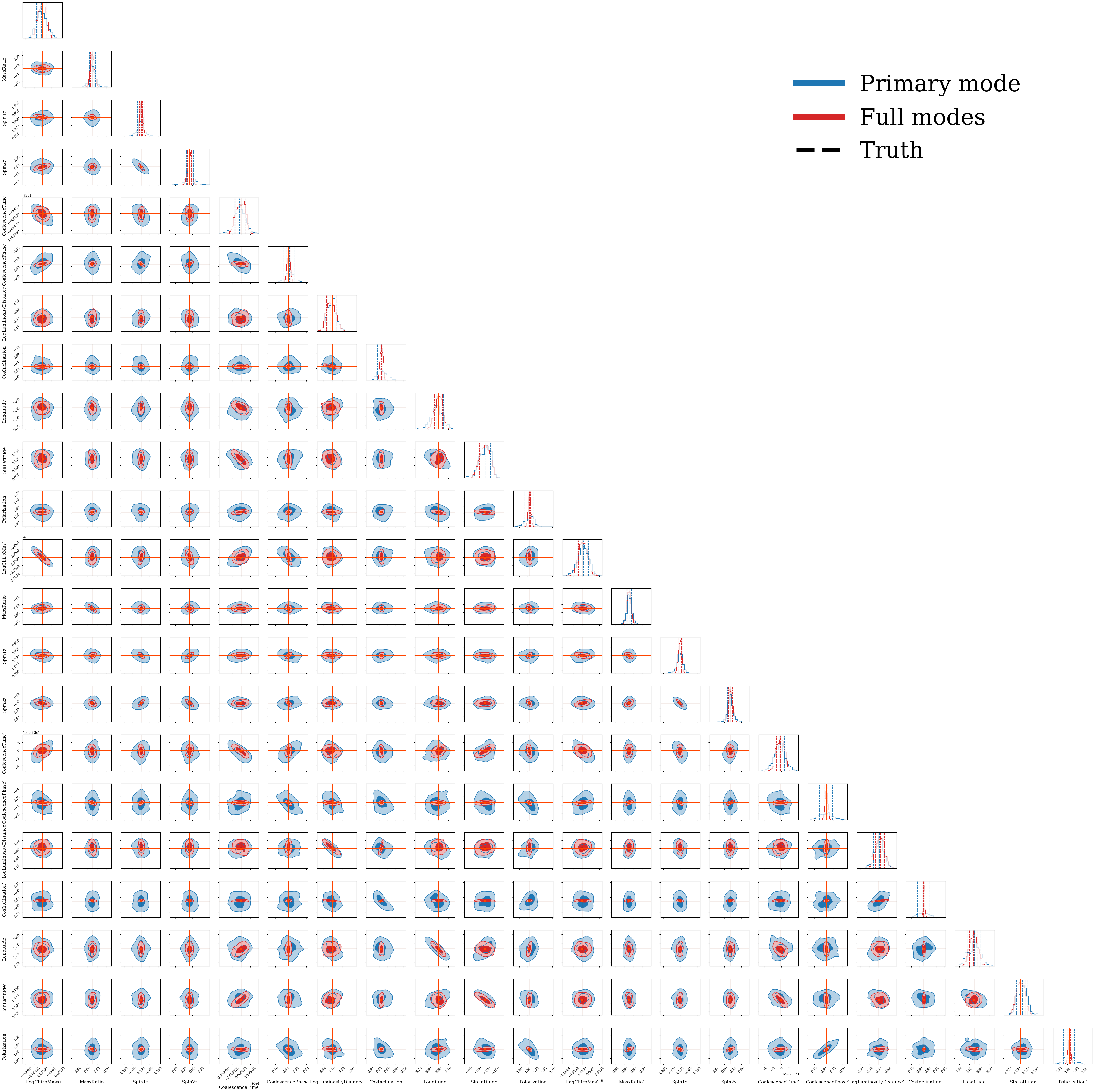}
\caption{Corner plot of the posterior distribution for two MBHB signals using MCMC sampling. Red contours indicate results with HMs; blue contours correspond to the dominant $(2,2)$ mode only. The shaded regions represent the 1$\sigma$ and 2$\sigma$ credible intervals.}
\label{fig:two_corner}
\end{figure*}
To further illustrate the impact of HMs, Figure~\ref{fig:two_corner} presents corner plots of the joint MCMC posterior distributions for two overlapping signals. The inclusion of HMs clearly reduces the overall parameter uncertainties and alleviates several parameter degeneracies. In particular, HMs helps disentangle correlations between the coalescence phase and coalescence time, between sky latitude and longitude, and between the mass ratio and the component spins. Notably, the substantial improvement in breaking the degeneracy between luminosity distance and inclination angle, previously reported in single‐source analyses, now extends to the overlapping‐source case, indicating that HMs can effectively disentangle these parameters even under source confusion. Nonetheless, the improvements are generally less pronounced than in the single-signal scenario. For certain  parameters, such as luminosity distance ($D_L$) , coalescence time ($\Phi_c$), and latitude ($\beta$), the uncertainties remain relatively moderate.

In summary, 
the inclusion of HMs generally 
improves parameter estimation by reducing parameter correlations and improving the estimation accuracy of intrinsic source parameters.  Furthermore, by incorporating HMs for accurate waveform modeling, the full Bayesian inference presented here corroborates the reliability of our Fisher‐based assessment of source confusion effects. 

\section{Conclusions}
\label{sec:conclusion}
In this study, we investigated the source confusion problem of MBHBs for the Taiji mission. 
Using the \textsc{PopIII}, \textsc{Q3d}, and \textsc{Q3nod} population models, we simulated MBHB populations and estimated the prevalence of signal overlap. 
Assuming a moderate local merger rate that yields roughly 100 events over one year of observation, we find that ``genuine'' source confusion, defined as overlap in both time and frequency domain, is relatively rare, with only 0.31 to 4.2 overlapping events per year on average depending on the population model. 
Therefore, it is practically sufficient to consider only the overlap between two signals. 

To quantify the impact on parameter inference, the uncertainty ratio $\gamma^i$ is defined within the framework of FIM. 
When GW signals exhibit similar time-frequency tracks, parameter uncertainties grow significantly. 
Thus, to establish an upper bound on the confusion effect, we performed Fisher matrix analyses across a broad range of chirp masses ($10^4$–$10^6~M_\odot$) with $\Delta \mathcal{M}_z = 0$. 
We find that the uncertainty ratio $\gamma^i-1$ increases with chirp mass and can reach $\sim10^2$ in the high-mass regime when only the dominant $(2,2)$ mode is used. 
The inclusion of HMs effectively mitigates this growth; even for nearly identical signals, uncertainty increases remain bounded within $\mathcal{O}(10^1)$ for most parameters relative to the individual-signal case.

Building upon this framework, we further investigated how uncertainty ratio depends on the chirp mass difference $\Delta\mathcal{M}_z$ and the phase difference $\Delta\Phi(f_{\mathrm{ov}})$. 
For variations in $\Delta\mathcal{M}_z$, the uncertainty ratio $\gamma^i - 1$ exceeds $\mathcal{O}(10^1)$ when the relative difference in chirp mass $\Delta \mathcal{M}_z / \mathcal{M}_z$ falls below 0.2\% for most parameters with HMs included. 
Such strong degeneracies are rare in MBHB populations, occurring in fewer than 0.14\% of cases under three population models. These findings suggest that severe source confusion can typically be neglected in practical MBHB data analysis. 
For variations in $\Delta\Phi(f_{\mathrm{ov}})$, $\gamma^i - 1$ reach their maximum when the two signals are phase-aligned ($\Delta\Phi \simeq 0$ or $\pi$) and minimize near $\Delta\Phi \simeq \pi/2$. The inclusion of HMs confines $\gamma^i - 1$ to within $\mathcal{O}(10^1)$ for the chirp mass and mass ratio, even when the two signals have identical chirp masses. 

Extending our Fisher-based analysis to more realistic data analysis scenario, we performed full Bayesian inference using MCMC for representative overlapping configurations exhibiting large $\gamma^i-1$ values. 
Crucially, HMs play a key role in breaking parameter degeneracies, particularly between the ecliptic latitude and longitude, thereby enabling more accurate sky localization and supporting effective multi-messenger follow-up observations. 
These enhancements are essential for accurate source characterization.

In summary, our results, validated by both Fisher forecasts and MCMC inference, demonstrate that while source confusion can degrade parameter estimation for MBHBs, its overall impact can be mitigated by incorporating HMs into waveform modeling and therefore be generally moderate. 
The Fisher matrix results provide a theoretical  bound on expected uncertainties and offer practical guidance for the development of data analysis strategies in future space-based GW missions. 

Looking ahead, future work may extend this framework to more complex scenarios involving the overlap among different source types and explore advanced inference techniques that further leverage waveform modeling improvements to enhance parameter recovery accuracy.
Furthermore, an implicit assumption of this work is that the number of MBHBs in the data is known a priori. 
The results of Fisher and MCMC analyses are based on matched filtering using known number of templates. 
For realistic data analysis in the future, it will be necessary to employ trans-dimensional search~\cite{10.1093/biomet/82.4.711,transMBHB,Future_trans_MBHB}  or iterative subtraction~\cite{Strub:2024kbe} techniques specifically to estimate unknown numbers of MBHBs.

\begin{acknowledgments}
This work is supported by National Key Research and Development Program of China
(Grant No. 2021YFC2201901, No. 2021YFC2201903, No. 2024YFC2207300), 
International Partnership Program of the Chinese Academy of Sciences, Grant No. 025GJHZ2023106GC.
\end{acknowledgments}

\bibliographystyle{apsrev4-2}
\bibliography{ref}

\end{document}